\documentclass[12pt]{article}
\usepackage{amssymb,latexsym,epsfig,cite}

\begin{document}

\begin{center}
{\Large \bf Supersymmetrically transformed \\ [1ex]
periodic potentials}
\end{center}

\begin{center}
David J. Fern\'andez C. \\ [2ex]
{\small
Departamento de F\'{\i}sica, CINVESTAV-IPN \\ A.P. 14-740,
07000 M\'exico D.F., Mexico }
\end{center}  

\noindent {\footnotesize
{\bf Abstract.} 
The higher order supersymmetric partners of a stationary periodic
potential are studied. The transformation functions associated to the band
edges do not change the spectral structure. However, when the
transformation is implemented for factorization energies inside of the
forbidden bands, the final potential will have again the initial band
structure but it can have bound states encrusted into the gaps, giving
place to localized periodicity defects.}

\bigskip

\section{Introduction}

Nowadays there is a growing interest in constructing exactly and
quasi-exactly solvable potentials which could serve as models in various
physical situations (see e.g. \cite{crf01} and references therein). There
exist simple generation techniques, e.g., the supersymmetric quantum
mechanics (SUSY QM) and other equivalent procedures as the Darboux
transformation, the factorization method, and the intertwining technique
\cite{ms91,ju96,ba01,cks01}. These constructions have been oftenly applied
to Hamiltonians having discrete energy levels
\cite{mi84,fe84,abi84,ais93,aicd95,fno96,fe97,bs97,fgn98,fhm98,ro98,sa00,mnr00,cr00,nnr00}. 
However, there are few works involving more general boundary conditions,
e.g., on periodic potentials for which the physical solutions of the
Schr\"odinger equation have to be bounded, and hence the spectrum of the
Hamiltonian is composed of allowed energy bands separated by the spectral
gaps (see e.g. \cite{rs78}). Our goal in this paper is to show that the
SUSY techniques applied to periodic Hamiltonians can produce new solvable
potentials \cite{tr89,df98,ks99,fnn00,sa01,asty01}. We will see that when
the transformation functions are taken as unphysical eigenfunctions of the
initial Hamiltonian with factorization energies inside of the forbidden
bands, new periodic or asymptotically periodic potentials can be generated
\cite{fmrs1,fmrs2}. The SUSY periodic partners will be produced when Bloch
transformation functions $u(x)$ are used, while the non periodic case will
arise for the general $u(x)$ (not necessarily in Bloch form). As a
byproduct, we will identify an interesting set of Darboux invariant
potentials, i.e., those first order SUSY partners which become just a
displaced copy of the initial potential \cite{fmrs2}.

The organization of the paper is as follows. In the second section a brief
overview of the first and second order SUSY QM will be presented. Then,
some simple facts about systems involving periodic potential will be
discussed. The SUSY QM will be then studied as a tool to generate solvable
potentials (periodic or asymptotically periodic) from an initial periodic
Hamiltonian.  The mechanism will be applied to the Lam\'e potentials, and
the paper will end up with a discussion about the Darboux invariance. 

\section{Intertwining technique}

Let us consider the following relationship
\begin{eqnarray}
&& \tilde H B^\dagger = B^\dagger H,  \label{intertwining} \\
&& H  =  - \frac{d^2}{dx^2} + V(x), \label{initialh} \\
&& \tilde H  = - \frac{d^2}{dx^2} + \tilde V(x), \label{finalh}
\end{eqnarray}
in which the operator $B^\dagger$ intertwines the two Schr\"odinger
operators $H, \ \tilde H$. Hence, if $\psi_n$ is an eigenfunctions of $H$
with eigenvalue $E_n$, $H\psi_n= E_n\psi_n$, then $\tilde\psi_n\propto
B^\dagger\psi_n\neq 0$ is an eigenfunction of $\tilde H$ with the same
eigenvalue, $\tilde H\tilde\psi_n= E_n\tilde\psi_n$. 

In case that $B^\dagger$ is a first order differential operator
\begin{eqnarray}
&& B^\dagger = -\frac{d}{dx} +
\alpha(x,\epsilon), \label{primero}
\end{eqnarray}
the Eqs.(\ref{intertwining}-\ref{finalh}) lead to the standard
interrelations between $\alpha(x,\epsilon), \ V(x), \ \tilde V(x)$: 
\begin{eqnarray}
& \alpha'(x,\epsilon)+
\alpha^2(x,\epsilon) = V(x)- \epsilon, \label{riccati} \\
& \tilde V(x) = V(x) -  2 \alpha'(x,\epsilon).\label{finalv} 
\end{eqnarray}
The Darboux formulae are obtained by taking $\alpha = [\ln u(x)]'$: 
\begin{eqnarray}
& -u''(x)+ V(x)u(x) = \epsilon u(x), \label{schro} \\  
& \tilde V(x) = V(x) -  2 [\ln u(x)]''. \label{finalvsh} 
\end{eqnarray}

Thus, a new solvable potential $\tilde V(x)$ can be efficiently generated
from $V(x)$
if one is able to solve explicitly (\ref{riccati}) or (\ref{schro}) for a
certain
$\epsilon$, where $\epsilon$ is called factorization constant because $H$
and $\tilde H$ admit the following factorizations: 

\begin{eqnarray}
&& H = B B^\dagger + \epsilon, \label{hfactorized} \\ &&
\tilde H = B^\dagger B + \epsilon . \label{newhfactorized}
\end{eqnarray}
Notice that $u(x)$ have to be nodeless inside the domain of $x$ in order
to avoid the creation of extra singularities of $\tilde V(x)$ with respect
to $V(x)$. This immediately leads to the typical restriction in the first
order case, $\epsilon \leq E_0$, where $E_0$ is the ground state energy of
$H$. 

The eigenvalues $E_n$ of $H$ for which $B^\dagger \psi_n \neq 0$ belong to
the spectrum of $\tilde H$. The rest of Sp($\tilde H$) depends on the
kernel of $B$, $B\tilde\psi_\epsilon = 0$. The solution of this last
equation, $\tilde\psi_\epsilon \propto 1/u(x)$, is an eigenfunction of
$\tilde H$ with eigenvalue $\epsilon$. According to the normalizability of
the non-singular $\tilde\psi_\epsilon$, we observe three different cases: 

\begin{itemize}

\item If $\epsilon=E_0$ and $u(x) = \psi_0(x)$, it turns out that
$\tilde\psi_\epsilon$ is non-normalizable $\Rightarrow {\rm Sp}(\tilde H)
= \{E_n,n=1,2,\dots\}$ (the level $E_0$ is `deleted' in order to get
$\tilde V(x)$).

\item If $\epsilon<E_0$ and non-normalizable $\tilde\psi_\epsilon$ can be
found $\Rightarrow {\rm Sp}(\tilde H)  = \{E_n,n=0,1,\dots\}$ (the
strictly isospectral case).

\item If $\epsilon<E_0$ and normalizable $\tilde\psi_\epsilon$ can be
found $\Rightarrow {\rm Sp}(\tilde H) = \{\epsilon,E_n,n=0,1,\dots\}$ (a
level is `created' at $\epsilon$ in order to get $\tilde V(x)$). 

\end{itemize}

\bigskip

Suppose now that $B^\dagger$ is a second order differential operator: 
\begin{eqnarray} 
&& B^\dagger = \frac{d^2}{dx^2} + \beta(x)\frac{d}{dx} +
\gamma(x).  \label{second} 
\end{eqnarray} 
By using again (\ref{intertwining}-\ref{finalh}), a pair of equations
generalizing (\ref{riccati}-\ref{finalv}) are found \cite{ais93,aicd95}:
\begin{eqnarray}
&& \frac{\beta''}{2\beta}-\left(\frac{\beta'}{2\beta}\right)^2
-\beta' + \frac{\beta^2}{4} +  \left(\frac{\epsilon_1 -
\epsilon_2}{2\beta}\right)^2  + \frac{\epsilon_1 + \epsilon_2}{2}
=  V(x), \label{sususyeq} \\
&& \hskip3cm \tilde V(x) = V(x) + 2 \beta'(x). \label{sususypot} 
\end{eqnarray} 

The solutions of the non-linear second order differential equation
(\ref{sususyeq}) can be found either in terms of those of the Riccati
equation (\ref{riccati}), $\alpha(x, \epsilon_1), \ \alpha(x,
\epsilon_2)$, or of those of the Schr\"odinger equation (\ref{schro}),
$u_1(x), \ u_2(x)$, $\epsilon_1 \neq \epsilon_2$: 
\begin{equation} 
\beta(x) = \frac{\epsilon_1 - \epsilon_2}{\alpha(x,
\epsilon_1) - \alpha(x, \epsilon_2)} = - [\ln W(u_1,u_2)]' .
\end{equation} 

Contrasting with the first order SUSY, in which $u_1$ and $u_2$ should be
nodeless, in the second order case the Wronskian $W(u_1,u_2)$ has to be
free of zeros, although $u_1$ and $u_2$ could have nodes. Once again,
Sp($\tilde H$)  depends on the normalizability of the two eigenfunctions
of $\tilde H$ with eigenvalues $\epsilon_1, \ \epsilon_2$ which belong
to the
Kernel of $B$. Their explicit expressions are:
\begin{equation} 
\tilde\psi_{\epsilon_1}\propto \frac{u_2(x)}{W(u_1,u_2)},
\quad \tilde\psi_{\epsilon_2}\propto \frac{u_1(x)}{W(u_1,u_2)}. 
\end{equation} 
Different cases can be reported: 

\begin{itemize}

\item If $\epsilon_1 = E_i$, $u_1 = \psi_i$, $\epsilon_2 = E_{i+1}$, $u_2
= \psi_{i+1}$, the two $\tilde\psi_{\epsilon_1}$,
$\tilde\psi_{\epsilon_2}$ are non-normalizable $\Rightarrow {\rm
Sp}(\tilde H) = \{E_0,\dots, E_{i-1},$ $E_{i+2},\dots\}$.

\item If $\epsilon_1 < \epsilon_2 < E_0$ and one can find normalizable
$\tilde\psi_{\epsilon_1}$ , $\tilde\psi_{\epsilon_2}$ $\Rightarrow {\rm
Sp}(\tilde H) = \{\epsilon_1,\epsilon_2,E_n, \\ n=0,1,2,\dots\}$. 

\item If $E_i<\epsilon_1<\epsilon_2<E_{i+1}$ and one can find normalizable
$\tilde\psi_{\epsilon_1}$ , $\tilde\psi_{\epsilon_2}$ $\Rightarrow {\rm
Sp}(\tilde H) = \{E_0,\dots,E_i,\epsilon_1,\epsilon_2,E_{i+1},\dots\}$. 

\end{itemize}

The intertwining technique and the supersymmetric quantum mechanics are
closely related, namely, the standard SUSY algebra
\begin{equation}
[Q_i, H_{\rm ss}]=0, \quad
\{Q_i,Q_j\} = \delta_{ij} H_{\rm ss}, \quad i,j=1,2, \label{susyqm}
\end{equation}
is simply realized by identifying
\begin{eqnarray}
& Q_1 =\frac1{\sqrt{2}}(Q^\dagger + Q),
\quad Q_2 = \frac1{i\sqrt{2}}(Q^\dagger - Q), \nonumber \\ 
& Q =  \left(\matrix{ 0 & 0
\cr B & 0 }
\right), \quad Q^\dagger =
\left(\matrix{ 0 & B^\dagger \cr 0 & 0 } \right),  \nonumber \\ [0.25cm] &
H_{\rm ss} =
\left(\matrix{ B^\dagger B & 0 \cr 0 & B B^\dagger }
\right). \nonumber 
\end{eqnarray}
The first order SUSY QM arises if $B^\dagger$ is the first order operator
of (\ref{primero}), and thus $H_{\rm ss}$ is {\it linear} in the matrix
operator $H^p = \left(\matrix{ \tilde H & 0 \cr 0 & H } \right)$:
\begin{equation}
H_{\rm ss} = (H^p-\epsilon).
\end{equation}
On the other hand, the second order SUSY (SUSUSY) QM arises if $B^\dagger$
is the second order operator (\ref{second}). In this case $H_{\rm ss}$
becomes {\it quadratic} in $H^p$ \cite{ais93,aicd95}: 
\begin{equation}
H_{\rm ss} = (H^p-\epsilon_1)(H^p-\epsilon_2).
\end{equation}
In general, in the higher order SUSY QM $B^\dagger$ is an $n$-th order
differential operator, $n>1$; in this paper we consider just the cases of
first and second order.

\section{ Schr\"odinger equation with \\ periodic potentials}

For periodic potentials $V(x+T)=V(x)$ it is convenient to work with the
stationary Schr\"odinger equation in the matrix form: 
\begin{equation}
\frac{d}{dx} \left[ 
\begin{array}{c}
\psi \\ \psi'
\end{array}
\right] = \left[
\begin{array}{cc}
0 & 1 \\ V(x)-E & 0
\end{array}
\right]
\left[ \begin{array}{c}
\psi \\ \psi'
\end{array}
\right]. \label{schromatrix}
\end{equation}
There is a linear mapping `propagating' the solution from a fixed point
(let us say $x=0$) to an arbitrary point $x$: 
\begin{equation}
\left[ \begin{array}{c}
\psi(x) \\ \psi'(x)   
\end{array} \right] = 
b(x)\left[ \begin{array}{c}
\psi(0) \\ \psi'(0)   
\end{array}  
\right]. \label{trans}
\end{equation}
The $2\times 2$ symplectic matrix $b(x)$ is called {\it transfer matrix}. 
The general behaviour of $\psi$ and the spectrum of $H$ depend on the
eigenvalues of the Floquet matrix $b(T)$, which in turn are determined by
{\it the discriminant} of $b(T)$, $D=D(E) = {\rm Tr} [b(T)]$:
\begin{equation}\label{CC}
\beta ^2 - D\beta + 1 = 0 \quad \Rightarrow \quad
\beta_{\pm}=D/2\pm \sqrt {D^2/4-1}, \quad 
\beta_+ \beta_- =1.
\end{equation}
The {\it Bloch functions} are particular solutions of (\ref{schromatrix}) 
arising if $[\psi(0),\psi'(0)]^{T}$ is one of the eigenvectors of $b(T)$,
i.e.,
\begin{equation}
\psi(T) = \beta \psi(0), \quad \psi'(T) = \beta \psi'(0).  \label{bloch}
\end{equation}
According to the values of $D(E)$, three different physical behaviours are
observed: 

\begin{itemize}

\item $|D(E)| < 2$. It turns out that $\beta_\pm\in{\mathbb C}$,
$|\beta_\pm|=1$. This implies that any solution (\ref{trans}) is bounded,
and hence $E$ belongs to an allowed energy band. 

\item $|D(E)| = 2$. Both $\beta_+$ and $\beta_-$ become either $+1$ or
$-1$, and the associated Bloch functions are periodic or antiperiodic
respectively. The Floquet matrix $b(T)$ is degenerated, and the values of
$E$ for which $|D(E)| = 2$, denoted as
$$ 
E_0<E_1\le E_{1'}<\ldots <E_j\le E_{j'}<\ldots
$$
define the band edges (which belong also to the spectrum of $H$).

\item $|D(E)| > 2$. Now $\beta_\pm\in{\mathbb R}$, implying that the
solutions of (\ref{trans}) are unbounded. Hence, $E$ belongs to a
forbidden energy band. 

\end{itemize}

\section{Supersymmetrically transformed \\ periodic potentials}

It is straightforward to apply the SUSY techniques of section 2 to the
periodic potentials of section 3 in order to generate solvable potentials
from a given initial one.  Let us employ first as transformation functions
the periodic or antiperiodic Bloch functions associated to the band edges.
The results are the following: 

\begin{itemize} 

\item 1-SUSY using the `ground state' eigenfunction $\psi_0$
\cite{df98,ks99}. It turns out that $\tilde V(x)$ is non-singular and
periodic.  As $B^\dagger$ maps bounded eigenfunctions of $H$ into bounded
ones of $\tilde H$, unbounded into unbounded, etc. $\Rightarrow$
$\tilde H$ and $H$ have the same band structure.

\item SUSUSY employing the band edge eigenfunctions $\psi_j,$ $\psi_{j'}$
bounding the energy gap $(E_j,E_{j'})$ \cite{fnn00}. The Wronskian becomes
nodeless and periodic $\Rightarrow$ $\tilde V(x)$ is periodic. Once again
$B^\dagger$ transforms bounded eigenfunctions of $H$ into bounded ones of
$\tilde H$, etc. $\Rightarrow$ $H$ and $\tilde H$ have the same band
structure.

\end{itemize}

A generalization employing Bloch functions with $\epsilon$ inside of a
forbidden energy band can be easily implemented. We distinguish some
interesting cases:

\begin{itemize}

\item 1-SUSY using Bloch eigenfunctions for $\epsilon\in(-\infty,E_0)$
\cite{tr89}. Those eigenfunctions are nodeless, leading to a non-singular
periodic superpotential $\alpha(x)$ $\Rightarrow$ $\tilde H$ and $H$ have
the same band structure. 

\item SUSUSY employing two nodeless Bloch eigenfunctions associated to
$\epsilon_1,\epsilon_2$ which belong to $(-\infty,E_0)$
\cite{fmrs1,fmrs2}. It turns out that $\beta(x)$ is non-singular and
periodic $\Rightarrow \ \tilde H$ and $H$ have the same band structure.

\item SUSUSY using two Bloch eigenfunctions with nodes for
$\epsilon_1,\epsilon_2$ such that $E_j < \epsilon_1 < \epsilon_2 < E_{j'}$
\cite{fmrs1,fmrs2}.  Once again $\beta(x)$ becomes non-singular and
periodic $\Rightarrow \ \tilde H$ and $H$ have the same band structure. 

\end{itemize}

\begin{figure}[ht]
\centering \epsfig{file=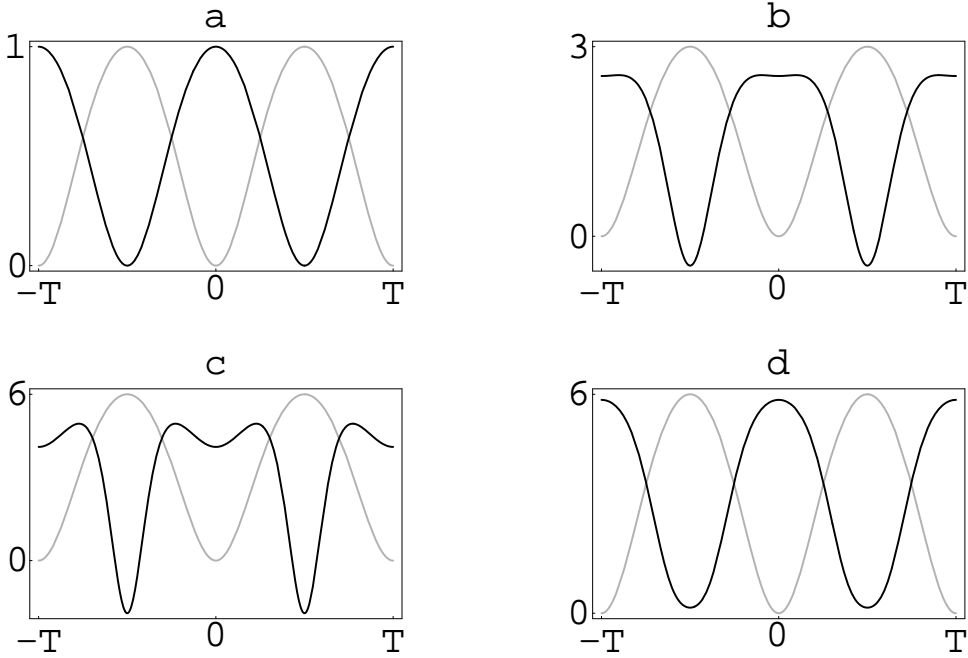, width=15cm}
\caption{\footnotesize The Lam\'e potentials (gray curves) of parameter
$m=1/2$ and their periodic SUSY partners (black curves) generated by
employing physical band edge eigenfunctions: a-c) 1-SUSY for
$\epsilon=E_0$ and $n=1,2,3$ respectively; d)  SUSUSY for $\epsilon_1=E_1,
\ \epsilon_2=E_{1'}$ and $n=3$.}
\end{figure}

Let us see next what happens when general (non-Bloch)  eigenfunctions of
$H$ are employed as transformation functions \cite{fmrs1,fmrs2}. 

\begin{itemize}

\item 1-SUSY using non-Bloch solutions of (\ref{schromatrix})  for
$\epsilon\in (-\infty,E_0)$. It turns out that nodeless $u$ can be found
such that $\tilde\psi_{\epsilon_1}$ is normalizable. The new potential
$\tilde V(x)$ has a local nonperiodicity but it is asymptotically
periodic. The operator $B^\dagger$ maps bounded eigenfunctions of $H$ into
bounded ones of $\tilde H$, unbounded into unbounded $\Rightarrow \ \tilde
H$ and $H$ have the same band spectrum but there is a bound state of
$\tilde H$ at $\epsilon$.

\item SUSUSY employing two non-Bloch eigenfunctions for
$\epsilon_1,\epsilon_2$ such that $\epsilon_1 < \epsilon_2 < E_{0}$.
$u_1,u_2$ can be selected such that $\beta$ is non-singular and
$\tilde\psi_{\epsilon_1}$, $\tilde\psi_{\epsilon_2}$ are normalizable.
$\tilde V(x)$ presents local nonperiodicities but it is asymptotically
periodic. As $B^\dagger$ maps bounded eigenfunctions of $H$ into bounded
ones of $\tilde H$, etc. $\Rightarrow$ $\tilde H$ and $H$ have the
same band structure but there are two bound states of $\tilde H$ at
$\epsilon_1, \epsilon_2$. 

\item A similar SUSUSY procedure with $\epsilon_1, \ \epsilon_2$ such that
$E_j<\epsilon_1<\epsilon_2<E_{j'}$ can `create' two bound states inside
the spectral gap $(E_j, E_{j'})$. 

\end{itemize}

A nice illustration of the SUSY techniques is given by the Lam\'e
potentials:
\begin{eqnarray}
& V(x) = n(n+1) m {\rm sn}^2(x|m), \quad n\in{\bf N},
\label{lame}
\end{eqnarray}
where ${\rm sn}(x|m)$ is a standard Jacobi elliptic function of parameter
$m\in[0,1]$. The spectrum of $H$ has $2n+1$ band edges defining $n+1$
allowed and $n+1$ forbidden energy bands. Some results applying the SUSY
techniques using the band edge eigenfunctions are shown in figure 1. 
Examples employing Bloch functions with factorization energies inside of
the forbidden energy bands arise in figure 2. Finally, some cases
implemented by means of non-Bloch transformation functions are illustrated
in figure 3. 

\begin{figure}[ht]
\centering \epsfig{file=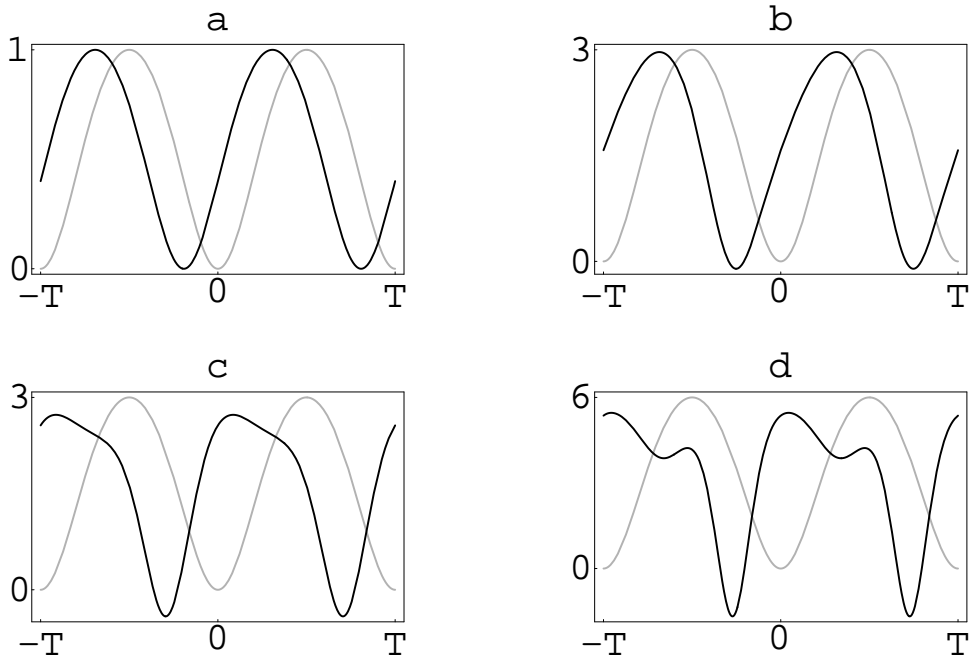, width=15cm} 
\caption{\footnotesize The Lam\'e potentials (gray curves) with $m=1/2$
and their periodic SUSY partners (black curves) generated by using
unphysical Bloch eigenfunctions: a) 1-SUSY with $\epsilon=-1, \ \epsilon <
E_0$ and $n=1$; b) 1-SUSY with $\epsilon=0.4, \ \epsilon<E_0$ and $n=2$;
c)  SUSUSY with $\epsilon_1=1.6, \ \epsilon_2=2.9$,
$E_1<\epsilon_1<\epsilon_2< E_{1'}$ and $n=2$; d) SUSUSY with
$\epsilon_1=2.3, \ \epsilon_2=5$, $E_1<\epsilon_1<\epsilon_2< E_{1'}$ and
$n=3$.}
\end{figure}

\begin{figure}[ht]
\centering \epsfig{file=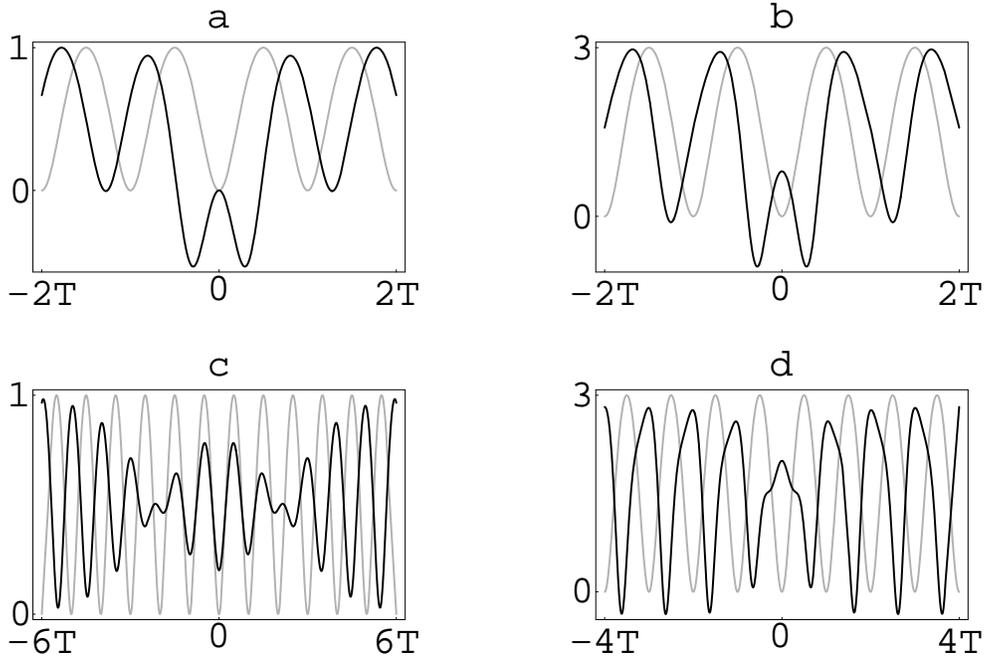, width=15cm}
\caption{\footnotesize The Lam\'e potentials (gray curves) with $m=1/2$
and their non-periodic SUSY partners (black curves) generated by using
unphysical non-Bloch eigenfunctions: a) 1-SUSY with $\epsilon=0, \
\epsilon < E_0$ and $n=1$; b) 1-SUSY with $\epsilon=0.4, \ \epsilon<E_0$
and $n=2$; c)  SUSUSY with $\epsilon_1=1.2, \ \epsilon_2=1.3$,
$E_1<\epsilon_1<\epsilon_2< E_{1'}$ and $n=1$; d)  SUSUSY with
$\epsilon_1=1.51, \ \epsilon_2=2.51$, $E_1<\epsilon_1<\epsilon_2< E_{1'}$
and $n=2$. All the non-periodic potentials have bound states at the
positions of the corresponding $\epsilon$'s.}
\end{figure}

Let us remark once again that when Bloch transformation functions are used
(see figures 1-2), the SUSY partner Hamiltonians $H$ and $\tilde H$ have
exactly the same spectrum. However, when non-Bloch solutions are employed
(see figure 3), the final potential `acquires' bound states encrusted into
the forbidden energy bands, which produces local non-periodicities of
$\tilde V(x)$. These potentials could be useful models for the contact
effects in solid state physics. 

\section{Darboux invariant potentials}

Let us notice that for the Lam\'e potential with $n=1$ the 1-SUSY
technique which employs the Bloch functions of either spectral gaps or
band edges produces a $\tilde V(x)$ which is just the initial potential of
a displaced argument, $\tilde V(x) = V(x+\delta)$ (see figures 1a and 2a). 
That phenomenon was discovered by Dunne and Feinberg for the lowest band
edge eigenfunction $u(x)=\psi_0(x)$ with $\delta = T/2$ (figure 1a), and
those potentials were called {\it selfisospectral} \cite{df98} (see also
\cite{fnn00}). Here we observe a more general invariance arising for $n=1$
and $\delta$ arbitrary, which is illustrated in figure 2a.  We propose the
name translationally invariant under Darboux transformation or simply {\it
Darboux invariant potentials} \cite{fmrs1,fmrs2}. It would be interesting
to seek when the 1-SUSY techniques induce that kind of symmetry. The
necessary and sufficient condition in order that the Bloch transformation
function $u^\beta(x)$ will produce a Darboux invariant potential is
\cite{fnn00,fmrs1}
\begin{equation} 
u^\beta(x)u^{1/\beta}(x+\delta) = {\rm constant}, \label{invariance}
\end{equation}
where $u^{1/\beta}(x)$ is the second Bloch eigenfunctions associated to
the factorization energy $\epsilon$. The restriction (\ref{invariance}) is
satisfied by the Bloch solutions for the Lam\'e potentials (\ref{lame}) 
with $n=1$ but it is not for $n>1$. We should be able to translate the
requirement (\ref{invariance}) into a restriction onto the form of the
potential which is going to be Darboux invariant. By using carefully the
1-SUSY techniques it can be shown that the Weierstrass potentials are the
only Darboux invariant potentials \cite{fmrs2}. In particular, the Lam\'e
potentials with $n=1$ are included in the Weierstrass family, but there
are inside also other interesting non-periodic ones as the 1-soliton well.
This result explains why the Lam\'e potential with $n=1$ is Darboux
invariant but those with $n>1$ are not. It has as well shed some light
about the general Darboux invariant potentials, not necessarily periodic.

\medskip

\noindent{\bf Acknowledgments.} The author acknowledges the support of
CONACYT, project 32086-E.

\end{document}